\documentclass{article}
\usepackage{t1enc}
\usepackage[utf8]{inputenc}
\usepackage[english]{babel}
\usepackage{graphicx}
\usepackage{amssymb}
\usepackage{amsmath}
\usepackage{amsthm}
\usepackage{enumerate}

\def\OA{{\cal A}}

\def\HS{{\cal H}}
\def\BO{{\cal B}}
\def\LO{{\cal L}}

\def\CO{\mathbb{C}}
\def\RE{\mathbb{R}}

\def\NA{\mathbb{N}}
\def\UN{\mathbf{1}}

\def\qed{\ \vrule height 5pt width 5pt depth 0pt}
\def\cros{\raise1.9pt\hbox{$\scriptscriptstyle
          >$}\!\raise1.5pt\hbox{$\scriptstyle\triangleleft\,$}}
\theoremstyle{definition}
\theoremstyle{definition}\newtheorem{Theo}{Theorem}
\theoremstyle{definition}\newtheorem{Prop}{Proposition}
\theoremstyle{definition}

\title{\bf An effective toy model in $M_n(\CO)$ for selective measurements in quantum mechanics}
\author{P\'eter Vecserny\'es\\ \\
\textit{Wigner Research Centre for Physics, Budapest, Hungary}
\thanks{email: vecsernyes.peter@wigner.mta.hu}}
\date{ }

\begin{document}
\maketitle

\begin{abstract}
The non-selective and selective measurements of a self-adjoint observable $A$ in quantum mechanics are interpreted as `jumps' of the state of the measured system into a decohered or pure state, respectively, characterized by the spectral projections of $A$. However, one may try to describe the measurement results as asymptotic states of a dynamical process, where the non-unitarity of time evolution arises as an effective description of the interaction of the measured system with the measuring device. 
The dynamics we present is a two-step dynamics: the first step is the non-selective measurement or decoherence, which is known to be described by the linear, deterministic Lindblad equation. 
The second step is a process from the resulted decohered state to a pure state, which is described by an effective non-linear `randomly chosen' toy model dynamics: the pure states arise as asymptotic fixed points, and their emergent probabilities are the relative volumes of their attractor regions. 
\vspace{0.1in}

\end{abstract}

\section{Introduction}\label{Intro}

In quantum mechanics a \emph{selective measurement} \cite{WZ} of a physical quantity $A$ (described by the self-adjoint operator $A\in\LO(\HS)$ on the Hilbert space $\HS$) leads to the following result: If $A=\int_{\sigma(A)} a\, dE_a$ is the spectral decomposition of $A$ and the system is described by the normal state $\omega\colon\BO(\HS)\to\CO$ then the outcome of the spectral interval $[a_1,a_2]\subset \sigma(a)$ occurs with probability (relative frequency) $\omega(E_{a_2}-E_{a_1})$ in the repeated experiments. During a single selective measurement with this result (i.e. $\omega(E_{a_2}-E_{a_1})\not= 0)$ the state of the system `jumps' into the normal state 
$\omega_{[a_1,a_2]}:=\omega\circ\Phi_{[a_1,a_2]}\colon\BO(\HS)\to\CO$ where
\begin{equation}\label{SM_on_observables}
\Phi_{[a_1,a_2]}(B):=\frac{(E_{a_2}-E_{a_1})B(E_{a_2}-E_{a_1})}{\omega(E_{a_2}-E_{a_1})},\quad B\in\BO(\HS).
\end{equation}
If $\HS$ is a finite $n$-dimensional Hilbert space then every state on $\BO(\HS)\simeq M_n(\CO)\equiv M_n$ is normal, 
i.e. can be uniquely given in terms of a density matrix $\rho\in \BO(\HS)_{+1}$, that is by a positive, trace one element in $\BO(\HS)$,  and the trace functional: $\omega( - )=\mathrm{Tr}\,(\rho - )$. The spectral decomposition of a self-adjoint operator $A\in  M_n$ can be written as a finite sum $A=\sum_{a\in\sigma(a)} a P_a$ in terms of commuting orthogonal spectral projections $\{ P_a\}$ that linearly span the unital abelian subalgebra 
$\langle A\rangle\subset M_n$ generated by $A$. In this case the result of a selective measurement is the state $\omega_a:=\omega\circ\Phi_a$ with probability $\omega(P_a)$, where
\begin{equation}\label{SM_on_observables_findim}
\Phi_a(B):=\frac{P_aBP_a}{\omega(P_a)},\quad B\in M_n.
\end{equation}
$\langle A\rangle$ is a maximal abelian subalgebra in 
$M_n$ iff the spectrum of $A$ is non-degenerate, i.e. only minimal orthogonal projections occur in the spectral decomposition of $A$. In this case the resulted state $\omega_a$ on $M_n$ is always pure.

For example, selective measurements of observables with infinite or finite dimensional abelian algebras $\langle A\rangle$ are the position measurement of an electron on a screen in a double-slit experiment, or the measurement of the spin component of an electron along a chosen axis in the Stern--Gerlach experiment, respectively.

A \emph{non-selective measurement} of a self-adjoint observable 
$A=\sum_{a\in\sigma(a)} a P_a \in M_n$ leads to a `re-preparation' of the original state into an $A$-decohered one: the original state 
$\omega$ `jumps' into the state 
$\omega_A:=\omega\circ \Phi_A$, where the map $\Phi_A\colon M_n\to \langle A\rangle'$ is the conditional expectation 
\begin{equation}\label{A_cond_exp}
\Phi_A(B):=\sum_{a\in\sigma(A)}P_a BP_a,\quad B\in M_n
\end{equation}
onto the commutant $\langle A\rangle'\subset M_n$ of $\langle A\rangle$. 
Since $\langle A\rangle$ is abelian it is contained in the image of $\Phi_A$: 
$\langle A\rangle\subseteq \langle A\rangle'$. Equality holds iff $\langle A\rangle$ is a maximal abelian subalgebra of $M_n$. 

It is a natural attempt to replace these `measurement jumps' of quantum states by a (very fast) dynamical process. However, there are objections to do this within the frame of usual time evolution in quantum theories. Although the $M_n\to M_n$ maps $\Phi_a$ and $\Phi_A$ in (\ref{SM_on_observables_findim}) and (\ref{A_cond_exp}), respectively, are completely positive (CP) maps ($\Phi_A$ is even unit preserving), they are not rank preserving in general. Therefore they destroy any unitary Heisenberg time evolution (even those with explicit time dependence) on the operators in $M_n$ because they cannot be written as $B\mapsto UBU^*$, i.e. by an adjoint map with a unitary $U\in M_n$.
Thus the question is whether these `measurement jumps' could be obtained as asymptotic states of a non-unitary (deterministic or stochastic) dynamical process. The non-unitarity of the underlying time evolution may arise as an effective description of the interaction with the measuring device or may be thought as the `true' fundamental dynamics of a quantum process (of measurements). 

The results of non-selective measurements are known to be described by asymptotic states of a  (deterministic, linear) Lindblad dynamics \cite{L} (see e.g. \cite{BaN}, \cite{W}). The generator of this non-unitary Heisenberg time evolution is the generator of a semigroup of unit preserving completely positive ($CP_1$) maps on $\BO(\HS)$. 

There are several dynamical models that describe the probabilistic outcomes of selective measurements or, in an other terminology, the collapse of wave functions. They use various (even gravity-related) non-linear stochastic dynamics \cite{GRW}, \cite{Pea}, \cite{D1}, \cite{Gi1}, \cite{Pen} or non-linear deterministic dynamics parametrized by hidden variables \cite{BB}. (For recent developments, see e.g. \cite{B}, \cite{D2}, \cite{BH}, \cite{Gi2}.) 
 
In a recent lecture \cite{Ge1} Tamás Geszti (see also \cite{Ge2}) raised the possibility of a non-linear dynamics where the outcomes of selective measurements would arise as asymptotic fixed points, and their emergent probabilities would be the relative volumes of their attractor regions. We present here a simple effective model of such a dynamics on the convex set of density matrices in finite dimensional Hilbert spaces. 
The model is a two-step description of selective measurements of a self-adjoint observable $A\in M_n$, 
where the non-unitarity of time evolutions in both steps are thought to be effective descriptions of the interaction of the measured subsystem with the measuring device.

The first step is a $CP_1$ time evolution given by a linear deterministic Lindblad dynamics \cite{L}, which leads to $A$-decohered asymptotic density matrices. The second step is a non-linear dynamics on $A$-decohered density matrices. The possible emergence of an effective non-linear dynamics from a unitary one can be supported by recent results: the Gross--Pitaevskii non-linear one-particle effective dynamics \cite{Gr}, \cite{Pit} can be derived from the unitary time evolution of the Bose-Einstein condensation if the number of particles tends to infinity \cite{LS}, \cite{ESY}. 
Here the prescribed non-linear deterministic dynamics leads to an $A$-pure density matrix from the $A$-decohered one. However, this dynamics contains a `randomly chosen' initial parameter, namely, an `external' or `trial' $A$-decohered density matrix 
$\rho_{ext}$. 
The choice of $\rho_{ext}$, that is the choice of the second step effective dynamics, is thought to reflect the (unknown) initial state of the full system (measured system and the measuring device) within the inverse image of the prepared initial state $\rho_0$ of the measured system, because any effective dynamics on the measured subsystem depends on the state of the full system due to the presence of the interaction with the measuring device. 
Fixing $\rho_{ext}$ the second step non-linear deterministic dynamics leads to an asymptotic fixed point, which is an $A$-pure density matrix $P_a, a\in\sigma(A)$, that is a spectral projection of the measured observable $A$. The non-linear dynamics reproduces the Born rule: repeated `experiments' with identically prepared initial state 
$\rho_0$ of the measured system but with a random choice of 
$\rho_{ext}$ from the uniformly distributed external density matrices lead to the probability 
(relative frequency) $\mathrm{Tr}\,(\rho_0P_a)$ of the possible asymptotic states $P_a, a\in\sigma(A)$.    

The outline of the paper is as follows. In Chapter 2 we briefly describe two types of effective non-unitary dynamics known in quantum theory, which have motivated the two different dynamics
used in our two-step effective description of selective measurements. First we discuss the effective dynamics given by a semigroup of $CP_1$-maps, and present the general form of the generator, called Lindblad generator. Then we cite the results how the non-linear Gross--Pitaevskii dynamics arises in a rigorous way as the one-particle effective description of Bose-Einstein condensation when the number of particles tends to infinity. 
In Chapter 3 we present our two-step toy model for selective measurements of a self-adjoint observable $A\in M_n$. 
For completeness we prove a proposition, based on known results,  about the necessary and sufficient conditions on asymptotic decoherence of density matrices in a Lindblad dynamics. 
Then we present our $\rho_{ext}$-dependent non-linear deterministic effective dynamics on decohered density matrices and prove a theorem: the stable fixed points of this dynamics are $A$-pure,
the measures of their attractor regions with uniformly distributed $\rho_{ext}$ are equal to the expectation values of the corresponding fixed point spectral projections of $A$ in the initial state of the first step Lindblad dynamics. Chapter 4 contains some closing remarks.

\section{Two types of effective dynamics in quantum theory}\label{2_eff_dynamics}

Since we do not want to modify the fundamental unitary dynamics of quantum theories we have to look for effective non-unitary dynamics for the description of selective measurements that arise from restrictions of unitary dynamics on (infinitely) large system to small (finite) subsystems. Examples for effective non-unitary dynamics exist, of course. A large class of linear deterministic non-unitary dynamics is given by semigroups of unit preserving completely positive ($CP_1$) maps. The second type of non-unitary dynamics we discuss shortly is a non-linear Schrödinger equation connected to a particular model: it is the one-particle effective non-linear dynamics of the Bose--Einstein condensation, the Gross--Pitaevskii dynamics.   

\subsection{The semigroup of unit preserving completely positive maps}

Completely positive maps have a natural relationship with subsystems in quantum theory. Already its definition
refers to the embeddings of the operator algebra $\BO(\HS)$ into 
$\BO(\HS)\otimes M_n,n\in\NA$ as a subsystem (= tensor product factor). The following two statements \cite{L} reinforce this relationship:

1. \textsl{If the group $U_t, t\in\RE$ of unitaries in 
$\BO(\HS_1)\otimes \BO(\HS_2)$ describe a Heisenberg time evolution on $\BO(\HS_1)\otimes\BO(\HS_2)$ then for any density matrix $\rho_2\in\BO(\HS_2)_{+1}$ the map
\begin{equation}\label{un_din_to_CP}
\BO(\HS_1)\ni A\mapsto \Phi_t(A):=\mathrm{Tr}_2\,
[(\UN_1\otimes\rho_2)
U_t^*(A\otimes \UN_2)U_t]\in\BO(\HS_1)
\end{equation}
is a $CP_1$ map for any time $t\in\RE$, where $\UN_i\in\BO(\HS_i)$ denotes the unit operator $i=1,2$.} 

Thus one can consider an effective $CP_1$ dynamics on the subsystem $\BO(\HS_1)$ instead of the unitary one. 
The `inverse' result is that any $CP_1$ map of a subsystem arises as a restriction of a unitary sandwiching of an extended system:

2. \textsl{If $\Phi$ is a $\sigma$-weakly continuous $CP_1$ map on $\BO(\HS_1)$ then there exists a Hilbert space $\HS_2$ and a $V$ isometry on $\HS_1\otimes \HS_2$ such that for any density matrix $\rho_2\in\BO(\HS_2)_{+1}$ the following equality holds:
\begin{equation}\label{CP_to_isometric_sandwiching}
\Phi(A)=\mathrm{Tr}_2\,
[(\UN_1\otimes\rho_2)
V^*(A\otimes \UN_2)V],\ A\in\BO(\HS_1).
\end{equation}
($V$ can be extended to a unitary element by allowing a $\rho_2$-dependent extension of $\HS_2$.)}  

Of course the maps $\Phi_t, t\geq 0$ in (\ref{un_din_to_CP}) do not form a semigroup, $\Phi_{t+s}\not=\Phi_t\circ\Phi_s$, in general. It is a further, Markovian type assumption that an effective dynamics can be described by a semigroup of $CP_1$ maps \cite{GKS}, \cite{L}. An important benefit of this assumption is that the generator $L$ of this semigroup can be `completely' characterized: \textsl{Let $L\colon\BO(\HS)\to\BO(\HS)$ be a bounded map with $L(\UN)=0$. 
Then $\exp(tL),t\geq 0$ are unit preserving $\sigma$-weakly continuous CP maps iff $L$ is given by
\begin{equation}\label{Lindblad_generator_H}
L(B)=i[H,B]+\sum_k (V_k^*BV_k -\frac{1}{2} \{V_k^*V_k,B\}),\ B\in\BO(\HS),
\end{equation}
where $H=H^*, V_k, \sum_k V_k^*V_k\in\BO(\HS)$ and $\{\ ,\ \}$ denotes the anti-commutator.} 

Clearly, this Heisenberg type time evolution can be translated to a Schrödinger type of evolution of a given state $\omega$ on 
$\BO(\HS)$ by defining the generator 
$\hat L(\omega):=\omega\circ L$. In case of a normal state 
$\omega$ given by the density matrix $\rho\in\BO(\HS)_{+1}$ this leads to the Lindblad equation 
\begin{equation}\label{Lindblad_equation}
\frac{d\rho}{dt}=\hat L(\rho):= -i[H,\rho]+\sum_k (V_k\rho V_k^* -\frac{1}{2} \{V_k^*V_k,\rho\}),
\end{equation}
which generalizes the Schrödinger equation containing only the self-adjoint Hamiltonian $H$ in the right hand side of (\ref{Lindblad_equation}). The property $L(\UN)=0$ implies unit preserving property of the maps $\Phi_t=\exp(tL), t\geq 0$, which is translated to the trace preserving property of the maps $\hat\Phi_t:=\exp(t\hat L)$, which remain $CP$ on $\BO(\HS)$ but they are not necessarily unit preserving.

It is known that the results of non-selective measurements can be described as asymptotic states of a Lindblad dynamics (\ref{Lindblad_equation}) by suitably chosen Lindblad operators $\{ H, V_k\}$, which is described in the next chapter.

\subsection{The effective non-linear Gross--Pitaevskii dynamics}

The trapped interacting $N$-boson Hamiltonian in three dimensions is given by
\begin{equation}\label{trapped_BE_Ham}
H_N^{trap}=\sum_{j=1}^N(-\Delta_{\mathbf{r}_j}+V_{ext}(\mathbf{r}_j))+
\sum_{i<j}^N V_N(\mathbf{r}_i-\mathbf{r}_j)
\end{equation}
on the symmetrized $N$-fold tensor product Hilbert space 
$\HS_S^{\otimes N}$ with $\HS:=L^2(\RE^3)$. The potential $V_{ext}$, which is responsible for trapping, has the property $0< V_{ext}(\mathbf{r})\to\infty, \vert\mathbf{r}\vert\to\infty$. The pair interactions are described by the `$N$-rescaled' potential $0< V_N(\mathbf{r})=N^2V(N\mathbf{r})$, where $V$ is a spherically symmetric, positive, compactly supported, smooth potential with scattering length $a_0$.

The conjectured effective one-particle description \cite{Gr}, \cite{Pit} is given by the non-linear Gross--Pitaevskii equation and energy functional $E_{GP}$ on the one-particle Hilbert space $\HS$:
\begin{eqnarray}\label{GP_equation}
i\partial_t\varphi(t)&=&-\Delta\varphi(t)+8\pi a_0
\vert\varphi(t)\vert^2\varphi(t),\quad \varphi(t)\in\HS, \Vert\varphi\Vert=1\\
\label{GP_energy}
E_{GP}(\varphi)&:=&\int d^3r(\vert\nabla\varphi(\mathbf{r})\vert^2
+V_{ext}(\mathbf{r})\vert\varphi(\mathbf{r})\vert^2
+4\pi a_0\vert\varphi(\mathbf{r})\vert^4),\quad \Vert\varphi\Vert=1.
\end{eqnarray}
Bose--Einstein condensation in the ground state of 
$H^{trap}_N$ (\ref{trapped_BE_Ham}) was proved by Lieb and Seiringer \cite{LS}. Their result reveals the precise connection to the effective GP description: 
\emph{Let $\psi_N$ be the ground state of $H^{trap}_N$ and let 
$\gamma^{(n)}_N, 1\leq n\leq N$ be its $n$-particle reduced density matrix. Let $\varphi_{GP}$ be the minimizer of the GP energy functional (\ref{GP_energy}). Then for any fixed $n$}
\begin{equation}\label{BE_cond_to_GP}
\lim_{N\to\infty}\gamma^{(n)}_N = \vert\varphi_{GP}\rangle\langle\varphi_{GP}\vert^{n\otimes}.
\end{equation}

One can ask what happens with the BE-condensed state $\psi_N$ if the trap is removed, that is when the evolution of the system is described by the Hamiltonian
\begin{equation}\label{BE_Ham}
H_N=\sum_{j=1}^N -\Delta_{\mathbf{r}_j}+
\sum_{i<j}^N V_N(\mathbf{r}_i-\mathbf{r}_j),
\end{equation}
An exact connection between the unitary dynamics of the 
BE-condensation and the effective non-linear (hence, non-unitary) one-particle GP-dynamics was given by Erd\H os, Schlein and Yau \cite{ESY}: 
\emph{Let $\psi_N(t)$ be the solution of the Schrödinger equation 
$i\partial_t\psi_N(t)=H_N\psi_N(t)$ with initial condition 
$\psi_N(0):=\psi_N$
and let $\gamma^{(1)}_N(t)$ be its one-particle reduced density matrix. Let $\varphi(t)$ be the solution of the GP-equation (\ref{GP_equation}) with initial condition $\varphi(0):=\varphi_{GP}$. Then for any $t\geq 0$ 
\begin{equation}\label{BE_dyn_to_GP_dyn}
\gamma^{(1)}_N(t)\to \vert\varphi(t)\rangle\langle\varphi(t)\vert,\quad N\to\infty
\end{equation}
pointwise for compact operators on $\HS$.}

This important result allows us to conclude that a given unitary dynamics on a full system (thought to be the measured subsystem plus the measuring device) may lead to a non-linear effective dynamics on a small subsystem (thought to be the measured subsystem, i.e. the abelian subalgebra generated by the measured self-adjoint observable) if the full system is `large enough' compared to the subsystem.

\subsection{Dependence of the effective dynamics on the initial state of the full system}

In this subsection we would like to highlight the fact that the emerging effective (CP or non-linear) dynamics on the subsystem launched from identical initial states may depend on the possible (different) initial states of the full system due to the presence of interactions.

The dependence of effective CP-dynamics on the initial density matrix of the full system within the inverse image of the initial density matrix of the subsystem is clear: Consider the CP-dynamics (\ref{un_din_to_CP}) in the Schrödinger picture
\begin{equation}\label{eff_CP_Sch}
\hat\Phi_t(\rho_{12}\vert\rho_1):=\mathrm{Tr}_2\, [U_t\rho_{12}U_t^*]\in\BO(\HS_1)_{+1},\quad t\geq 0,
\end{equation}
where $\rho_1=\mathrm{Tr}_2\, \rho_{12}$ is the 
initial density matrix in the subsystem $\BO(\HS_1)$ and 
$\rho_{12}$ is one of the initial density matrices in the full system $\BO(\HS_1\otimes\HS_2)$ within the inverse image of $\rho_1$
\begin{equation}
\mathrm{Tr}_2^{-1}(\rho_1):=\{ \rho\in\BO(\HS_1\otimes\HS_2)_{+1}\,\vert\, 
\mathrm{Tr}_2\,(\rho)=\rho_1\}.
\end{equation}
Clearly, the effective time evolution (\ref{eff_CP_Sch}) of $\rho_1$ heavily depends on the choice of the initial density matrix from
$\mathrm{Tr}_2^{-1}(\rho_1)$ if $U_t$ is different on the `$\BO(\HS_1)$-blocks' within $\BO(\HS_1\otimes\HS_2)$, that is if $U_t$ is not a product of unitary elements from $\BO(\HS_1)$ and $\BO(\HS_2)$.

The initial state dependence of the effective non-linear one-particle dynamics in BE-condensation is less obvious due to the large $N$ limit procedure involved. In the light of the two theorems about BE-condensation above one can say the following: The initial one-particle GP density matrix $\rho_{GP}:=\vert\varphi_{GP}\rangle\langle\varphi_{GP}\vert$ on $\HS$ leads to a sequence of inverse image sets $R_N^{-1}(\rho_{GP})$ of density matrices on 
$\HS_S^{\otimes N}, N\in\NA$, where $R_N$ denotes the one-particle reduced density matrix map from $\BO(\HS_S^{\otimes N})_{+1}$ to $\BO(\HS)_{+1}$. Since $\gamma_N^{(1)}:=
R_N(\vert\psi_N\rangle\langle\psi_N\vert)$ (\ref{BE_cond_to_GP}) implies that the sequence of ground state density matrices $\vert\psi_N\rangle\langle\psi_N\vert$ of $H^{trap}_N$ on $\HS_S^{\otimes N}, N\in\NA$ approaches the inverse image set sequence $R_N^{-1}(\rho_{GP}), N\in\NA$ for large $N$. Moreover, choosing $\vert\psi_N\rangle\langle\psi_N\vert, N\in\NA$ as a particular initial state sequence with unitary dynamics $H_N$ (\ref{BE_Ham}) on $\HS_S^{\otimes N}, N\in\NA$  the non-linear GP-dynamics emerges in the large $N$ limit as the effective one-particle dynamics on $\HS$ due to (\ref{BE_dyn_to_GP_dyn}). 
Since any initial density matrix sequences from the sequence of inverse image sets $R_N^{-1}(\rho_{GP}), N\in\NA$ lead to a constant one-particle reduced density matrix sequence, namely 
$\rho_{GP}$, by definition, one conjectures that clever choices from such initial density matrix sequences (subject to the unitary dynamics $H_N, N\in\NA$ (\ref{BE_Ham})) lead to one-particle effective dynamics in the large $N$-limit different from the GP-dynamics. Although, there is no real hope to verify this guess we think that similarly to effective CP dynamics the emerging effective non-linear dynamics (involving large $N$-limits) depend on the choice of the initial state of the full system within the inverse image of the prescribed initial state of the subsystem.

Keeping this dependence in mind one can perform repeated runs of a given unitary dynamics on the full system with fixed initial condition 
$\rho_1\in\BO(\HS_1)_{+1}$ on the subsystem but with a random choice of initial condition $\rho\in \mathrm{Tr}_2^{-1}(\rho_1)$ on the full system with respect to a probability distribution on the inverse image set $\mathrm{Tr}_2^{-1}(\rho_1)$. According to the considerations above these repeated runs will lead to a probability distribution on the different effective dynamics of the subsystem through relative frequencies. Such considerations, that is non-uniqueness and probabilistic description of the emerging effective dynamics, will be taken into account in our effective toy model of selective measurements.

\section{A two-step effective dynamics for selective measurement in QM}\label{eff_time_evo_SM} 

Our toy model serves as an effective, two-step dynamical description of selective measurement of a self-adjoint observable $A\in M_n$. 
The two consecutive dynamical steps use the two types of effective dynamics on the set $S_n\equiv (M_n)_{+1}$ of density matrices discussed in the previous chapter: $CP_1$-dynamics and 
(a parametrized set of) non-linear dynamics. 

The first dynamical step, the Lindblad dynamics (\ref{Lindblad_equation}) on density matrices in $S_n$ is characterized by suitably chosen Lindblad operators, 
which results an $A$-decohered asymptotic density matrix $\rho_\infty\in S_A:=S_n\cap\langle A\rangle'$ from the initial one $\rho_0\in S_n$. The second dynamical step uses $\rho_\infty\in S_A$ as the initial state $\mu_0\equiv\rho_\infty$ of a non-linear effective dynamics on $S_A$. The possible non-linear effective dynamics are parametrized by $A$-decohered `external' density matrices, $\{\mu_{ext}\}=S_A$. 
The choice of the dynamics (parametrized by 
$\mu_{ext}\in S_A$) in a single-run measurement is thought to be the effective description of the choice of the initial state of the full system (measuring device plus measured subsystem) within the inverse image of the prepared initial state $\rho_0$ of the measured subsystem. We will prove that depending on the relative positions of $\mu_0$ and $\mu_{ext}$ the asymptotic state of the non-linear dynamics on $S_A$ is one of the spectral projection $P_a, a\in\sigma(A)$ of $A$, the relative frequency of the outcome $P_a$ in repeated experiments with identical initial states $\rho_0$ with respect to uniform distribution of 
$\mu_{ext}\in S_A$ is given by $\mathrm{Tr}\, (\rho_0P_a)$. 

\subsection{Decoherence due to specific Lindblad dynamics}

The possibility that certain Lindblad dynamics $\hat\Phi_t\equiv\exp(t\hat L), t\geq 0$ given by (\ref{Lindblad_equation}) could lead to asymptotic decoherence of an initial state with respect to the measured observable was known in the community of measurement theorists.
In case of a finite dimensional Hilbert space a detailed analysis of possible Lindblad evolutions of density matrices was given in \cite{BaN}. Recently, the decoherence of the initial density matrix in non-selective measurements was described in \cite{W} using Lindblad dynamics. For completeness, we formulate this decoherence in a proposition based on these earlier results. 

In the following we refer to a density matrix $\rho_\infty\in S_n$ as \emph{asymptotic state} of a Lindblad dynamics if there is an initial state $\rho\in S_n$ such that 
$\lim_{t\to\infty} \hat\Phi_t(\rho)=\rho_\infty$. Since the maps $\hat\Phi_t,t\geq 0$ form a semigroup, $\rho_\infty\in S_n$ is an asymptotic state iff it is an \emph{invariant state} with respect to the dynamics, $\hat\Phi_t(\rho_\infty)=\rho_\infty,t\geq 0$.

\Prop\label{A-decoherence} \textsl{i) The set of asymptotic states of a Lindblad evolution (\ref{Lindblad_equation}) with $\hat L=\hat L(H,V_k)$
is equal to the image $\Phi_A(S_n)$ of non-selective measurements (\ref{A_cond_exp}) of the self-adjoint observable 
$A\in M_n$ iff the von Neumann algebra generated by the Lindblad operators satisfies $\{H, V_k, V_k^*\}''=\langle A\rangle$.} 

\textsl{ii) Let $\{H, V_k, V_k^*\}''=\langle A\rangle$. Any initial state $\rho\in S_n$ leads to an asymptotic state iff 
$\{H, V_k, V_k^*\}''=\{V_k, V_k^*\}''$. Then}
\begin{equation}\label{form_as_state}
\lim_{t\to\infty} \hat\Phi_t(\rho)=\Phi_A(\rho) :=\sum_{a\in\sigma(A)}P_a\rho P_a.
\end{equation}

\noindent\emph{Proof.} i) We use the result in \cite{BaN} claiming that 
$B\in M_n$ is invariant, $\Phi_t(B)=B, t\geq 0$,  with respect to a Heisenberg $CP_1$-dynamics $\Phi_t:=\exp(tL), t\geq 0$ 
with Lindblad generator $L=L(H,V_k)$ (\ref{Lindblad_generator_H}) iff $B\in\{H, V_k, V_k^*\}'$.

Let the Lindblad operators be chosen in a way that $\{H, V_k, V_k^*\}''$ is equal to the commutative algebra 
$\langle A\rangle$ generated by the measured observable $A\in M_n$. Then $V_kV_k^*=V_k^*V_k$ for all $k$, hence 
$\hat L(H, V_k)$ given in (\ref{Lindblad_equation}) itself is a generator of a semigroup of $CP_1$ maps on $M_n$ given by (\ref{Lindblad_generator_H}) with 
$L(-H,V_k^*)=\hat L(H,V_k)$.
Applying the above mentioned result of \cite{BaN} to 
$\hat L$, the invariant subalgebra of $M_n$ with respect to 
$\hat\Phi_t,t\geq 0$ is equal to $\{ -H, V_k^*, V_k\}'=\langle A\rangle'=\Phi_A(M_n)$. Therefore the set of asymptotic states, that is the set of $\hat\Phi_t$-invariant density matrices is equal to 
$\{ -H, V_k^*, V_k\}'\cap S_n=\Phi_A(S_n)$.

For the opposite implication let us note first, that if the set of $\hat\Phi_t$-invariant states is $\Phi_A(S_n)$ 
then the $\hat\Phi_t$-invariant subalgebra of $M_n$ is 
$\Phi_A(M_n)$ because 
$\Phi_A$ is a linear map and $S_n$ linearly spans $M_n$. Since 
$\Phi_A$ is unit preserving the unit $\UN\in M_n$ is $\hat\Phi_t$-invariant, that is $\hat L(\UN)=0$, which implies $\sum_k V_kV_k^*=\sum_kV_k^*V_k$ due to the form (\ref{Lindblad_equation}) of $\hat L$. Thus $\hat L(H, V_k)$ is equal to the generator 
$L(-H,V_k^*)$ of a semigroup of $CP_1$ maps on $M_n$, therefore the $\hat\Phi_t$-invariant subalgebra, which is $\Phi_A(M_n)=\langle A\rangle'$ by assumption, is equal to $\{ -H, V_k^*, V_k\}'$ by \cite{BaN}. 
Hence, $\langle A\rangle=\{ H, V_k, V_k^*\}''$.

ii) The adjoint of the Lindblad generator $\hat L(H, V_k)$ in 
(\ref{Lindblad_equation}) with respect to the scalar product on $M_n$ given by the trace turns out to be the corresponding Lindblad generator (\ref{Lindblad_generator_H}), that is 
$\hat L(H, V_k)^*=L(H, V_k)$. Therefore $\hat L$ is not self-adjoint in general, but a Jordan decomposition of $\hat L$ as a linear map on $M_n$ exists. Thus the generalized eigenvalue problem for eigenmatrices of $\hat L$, that is the equation 
$(\hat L-\lambda)^k=0, k\geq 1$ for $k$-dimensional Jordan blocks spanned by $A_1,A_2,\dots, A_k\in M_n$ leads to the solution of the Lindblad equation (\ref{Lindblad_equation}): 
\begin{equation}\label{gen_sol_Lindblad_eq}
\hat\Phi_t(A_k)=e^{\lambda t} A_k,\  \hat\Phi_t(A_{k-1})=e^{\lambda t} (A_{k-1}+tA_k),\dots, \hat\Phi_t(A_1)=e^{\lambda t} (A_1+tA_2+\dots +t^{k-1}A_k).
\end{equation}
In our case $\{H, V_k, V_k^*\}''=\langle A\rangle$ by assumption. Commutativity of this algebra implies that $\hat L$ is normal, $\hat L^*\hat L=\hat L\hat L^*$, and 
$\hat L(H, V_k)=L(-H, V_k^*)$. Hence, only $k=1$ dimensional blocks occur in the Jordan decomposition of $\hat L$, and $\hat\Phi_t, t\geq 0$ themselves become $CP_1$ maps. However, a unit preserving positive map between $C^*$-algebras is a norm 1 map (see e.g. \cite{BR} Corollary 3.2.6.), 
that is 
$\Vert\hat\Phi_t(A)\Vert\leq\Vert A\Vert$ for all $t\geq 0$ and 
$A\in M_n$. Therefore $\mathrm{Re}\,\lambda\leq 0$ 
for (proper, $k=1$) eigenmatrices of $\hat L$ in (\ref{gen_sol_Lindblad_eq}), 
which implies that $\lim_{t\to\infty}\hat\Phi_t(A)=0$ for 
$\hat L$-eigenmatrices $A$ with $\mathrm{Re}\,\lambda < 0$, 
and the asymptotic states lie in the linear subspace of $M_n$ spanned by $\hat L$-eigenmatrices with 
$\mathrm{Re}\,\lambda = 0$. It has been shown in \cite{W} 
that the $\mathrm{Re}\,\lambda = 0$ eigensubspace is given by 
$\{ V_k, V_k^*\}'$. The restriction of $\hat L$ in (\ref{Lindblad_equation}) to this subspace becomes the anti-self-adjoint operator $\hat L(B)=-i[H,B]$, which has purely imaginary eigenvalues. Hence, every initial state $\rho\in S_n$ leads to an asymptotic state iff these purely imaginary eigenvalues are zero, that is iff $H\in\{ V_k, V_k^*\}''$, or equivalently $\{H, V_k, V_k^*\}''=\{V_k, V_k^*\}''$. The assumption
$\{H, V_k, V_k^*\}''=\langle A\rangle$ implies that 
$\Phi_A\circ \hat\Phi_t=\hat\Phi_t\circ\Phi_A, t\geq 0$ and 
since $\Phi_A$ being a norm one map is continuous (\ref{form_as_state}) also follows.\qed

\medskip
We close this subsection by emphasizing that an idealized two-step dynamical description of selective measurements  
is possible just because the first step, the non-selective 
measurement of $A$, preserves the initial expectation values within $\langle A\rangle'$, that is 
$\omega(C)=\omega_\infty(C):=\lim_{t\to\infty}\omega(\Phi_t(C))$ for $C\in\langle A\rangle'$. 
Indeed, due to (\ref{form_as_state}) we have 
$\omega_\infty(B)=\omega(\Phi_A(B)), B\in M_n$ and 
$\Phi_A(B)=B$ if $B\in\langle A\rangle'=\Phi_A(M_n)$ by using of (\ref{A_cond_exp}). 
Therefore the probability (relative frequency) of the spectral outcome $a\in\sigma(A)$ in the selective measurements with initial states 
$\omega$ and $\omega_\infty$ are equal, 
$\omega(P_a)=\omega_\infty(P_a)$, because 
$P_a\in\langle A\rangle\subseteq\langle A\rangle'$. Therefore 
the Born rule is not violated if 
the initial state of the second dynamical step, which is responsible for $A$-purification, will be the $A$-decohered asymptotic state $\omega_\infty=\omega\circ\Phi_A$ of the first.
In terms of density matrices this idealized two-step dynamics means that the initial density matrix $\mu$ of the second dynamical step will be the $A$-decohered asymptotic density matrix $\rho_\infty=\Phi_A(\rho)\in\Phi_A(S_n) 
=S_n\cap\langle A\rangle'$ of the first.

To be more precise, the second type dynamics will work on the set $S_A$ of density matrices on the abelian algebra $\langle A\rangle$. Clearly, $S_A=\Phi_A(S_n)$ only if $\langle A\rangle$ is a maximal abelian subalgebra in $M_n$, i.e. when $\langle A\rangle=\langle A\rangle'$. Hence, if $\langle A\rangle$ is not maximal abelian in $M_n$ then the initial density matrix $\mu\in S_A$ of the second dynamical step will be the restriction of the $A$-decohered density matrix $\rho_\infty=\Phi_A(\rho)$ to 
$\langle A\rangle$.

\subsection{Purification due to non-linear dynamics}

As we have seen in Chapter \ref{2_eff_dynamics} effective non-linear dynamics may arise from the restriction of a large $N$ limit unitary dynamics to a fixed small subsystem. We have also indicated that the effective dynamics depends on the initial state of the full system within the inverse image of the initial state of the subsystem.
The effective non-linear toy dynamics we present here is along these lines: The subsystem is the abelian algebra 
$\langle A\rangle$ generated by the measured observable $A$, 
the effective non-linear dynamics are given 
on the convex set $S_A:=\langle A\rangle_{+1}$ of density matrices on $\langle A\rangle$. The possible dynamics are characterized by external density matrices $\mu_{ext}\in S_A$ being fixed in a single run of the dynamics. The choice of the external density matrix $\mu_{ext}$ from $S_A$
is thought to be the effective description of the choice of the (unknown) initial state of the (unknown) full system (measuring device plus measured subsystem) within the inverse image of the initial state $\mu(0)\in S_A$ of the subsystem. 
Therefore $\mu_{ext}$ 
serves as an effective characterization of the dependence of the effective dynamics on the (unknown) initial state of the (unknown) full system. 

\begin{figure}[ht]
\centerline{\resizebox{6cm}{!}{\includegraphics{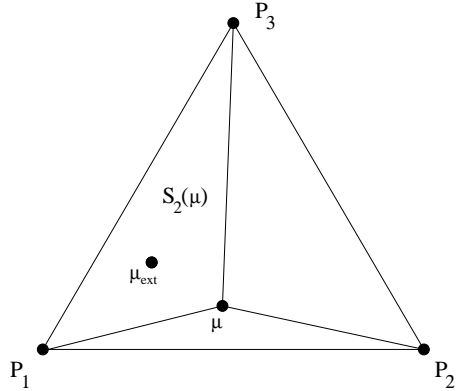}}}
\caption{The convex set $S_A$ of $A$-decohered density matrices in $\langle A\rangle\simeq \oplus^3 M_1$ is shown, which is spanned by its extremal points, the spectral projections $P_1, P_2, P_3$ of $A$. The subsimplices $S_i(\mu)$ are also indicated, but only 
$S_2(\mu)$ (spanned by $P_1, P_3$ and $\mu\in S_A$) is labeled, which contains the external density matrix $\mu_{ext}\in S_A$. Hence, it can be written as a convex combination: $\mu_{ext}=\lambda_2\mu+\lambda_1P_1+\lambda_3P_3\in S_2(\mu)$.}
\label{S_triangle}
\end{figure}

The non-linear dynamics on $S_A$ will be given by a first order non-linear differential equation characterized by a map 
$f\colon S_A\times S_A\to \langle A\rangle$.
Let $\mu, \mu_{ext}\in S_A\subset\langle A\rangle$ then the time evolution of $\mu$ is given by
\begin{equation}\label{nonlin_eff_dyn_SM}
\frac{d\mu}{dt}= f(\mu,\mu_{ext})-
\mu\,\mathrm{Tr}\, f(\mu,\mu_{ext}), \qquad
f(\mu,\mu_{ext}):=a\mu(\lambda \mu-\mu_{ext}),
\end{equation}
where $a>0$ is a constant and the real number $\lambda\equiv\lambda(\mu,\mu_{ext})$ as a function of $\mu$ and $\mu_{ext}$ is defined to be the maximal value of $\kappa\in[0,1]$ for which $\mu_{ext}-\kappa\mu$ is a positive operator. 
To clarify the value $\lambda$ let us note first that since 
$\langle A\rangle\simeq \oplus^n M_1$ $S_A$ is the convex hull of its extremal points $P_i, i=1,\dots, n$, which are the minimal projections in 
$\langle A\rangle$. Let the subsimplex $S_i(\mu)\subset S_A, i=1,\dots, n$ be defined as the convex hull of $P_1, \dots, P_{i-1}, \mu,P_{i+1},\dots, P_n$. Clearly, $\cup_i S_i(\mu)=S_A$. 
If $\mu_{ext}\in S_i(\mu)$ then 
it can be written as a convex combination $\mu_{ext}=\lambda_i\mu +\sum_{k\not=i}\lambda_kP_k$ and one arrives at 
$\lambda(\mu,\mu_{ext})=\lambda_i$. For a special case with $n=3$ see Figure \ref{S_triangle}.

Although the dynamics (\ref{nonlin_eff_dyn_SM}) is deterministic the relative frequency of the outcomes, that is the asymptotic states with identical initial conditions $\mu(0)$ will depend on the distribution of the choice of external density matrices 
$\mu_{ext}\in S_A$ in the repeated runs. Thus we have to define a measure on $S_A$, that is on the convex hull of the minimal projections in $P_i, i=1,\dots, n\in\langle A\rangle$. Using the trace as a scalar product on the real vector space $H_A$ of self-adjoint elements in $\langle A\rangle$, $H_A$ becomes a  real Hilbert space isomorphic
to $\RE^n$ through the mapping of the minimal projections 
$P_1,\dots, P_n\in\langle A\rangle$ into an orthonormal basis of $\RE^n$. Hence, $S_A\subset H_A\simeq \RE^n$ becomes a $(n-1)$-dimensional simplex in the unit cube of $\RE^n$, and the Lebesgue measure in $\RE^{n-1}$ provides a (finite) measure on $S_A$.

\begin{Theo} \textsl{Let $\langle A\rangle\subset M_n(\CO)$ be a  maximal abelian subalgebra generated by a self-adjoint element $A$. Let the real Hilbert space $H_A$ of self-adjoint elements in 
$\langle A\rangle $ be identified with $\RE^n$ through the orthonormal basis given by the minimal projections 
$P_1,\dots, P_n\in\langle A\rangle$.} 

\textsl{i) Let $\mu(0)\in S_A$ be an initial density matrix for the differential equation (\ref{nonlin_eff_dyn_SM}) with an arbitrary but fixed external density matrix $\mu_{ext}$ in the interior $S_A^{int}$ of the closed convex set $S_A\subset \RE^n$. Then there exists a unique solution of the initial value problem and the corresponding integral curve $\mu(t),t\geq 0$ lies in $S_A$. The generic asymptotic states for any pair $\mu(0)\in S_A$ and $\mu_{ext}\in S_A^{int}$ are the minimal projections $P_1,\dots, P_n\in\langle A\rangle$, which are the stable fixed points of the dynamics (\ref{nonlin_eff_dyn_SM}) on $S_A$.}

\textsl{ii) Let the measure on $S_A$ be induced by the Lebesgue measure in $\RE^{n-1}$ through the image of $S_A\subset H_A$ in $\RE^n$, which is a $(n-1)$-dimensional simplex in the unit cube of $\RE^n$. If the external density matrix $\mu_{ext}\in S_A$ is chosen uniformly random with respect to this measure then 
the asymptotic states of (\ref{nonlin_eff_dyn_SM}) started from identical initial states 
$\mu(0)$ are equal to $P_i$ with probability 
$\mathrm{Tr}\, (\mu(0) P_i), i=1,\dots, n$.}
\end{Theo}

\begin{figure}[ht]
\centerline{\resizebox{6cm}{!}{\includegraphics{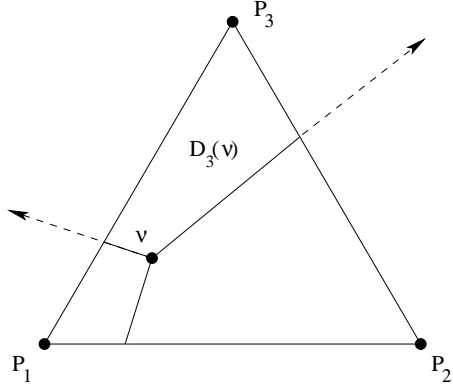}}}
\caption{In case of $\langle A\rangle\simeq \oplus^3 M_1$ the closed domains $D_i(\nu):=C_i(\nu)\cap S_A, i=1,2,3$ in the convex set $S_A=\langle P_1,P_2,P_3\rangle$ of $A$-decohered density matrices are shown. They are bounded by straight lines, the domain $D_3(\nu)$ is even labeled. The (two) extremal rays (generated by $\nu-P_1$ and $\nu-P_2$) of the affine convex cone $C_3(\nu)$ are indicated by dashed lines.}
\label{D_domains}
\end{figure}

\noindent \emph{Proof.} i) 
The Picard--Lindelöf theorem for first order differential equations provides the existence of a unique solution of the initial value problem in a region of $S_A$ if uniform Lipschitz continuity holds there for the tangent vector $F(\mu,\mu_{ext}):=f(\mu,\mu_{ext})-
\mu\,\mathrm{Tr}\, f(\mu,\mu_{ext})\in H_A\simeq\RE^n$, i.e. for the right hand side of (\ref{nonlin_eff_dyn_SM}). Hence, it is enough to prove uniform Lipschitz continuity 
\begin{equation}\label{Lip_cont}
\Vert(F(\mu,\mu_{ext})-F(\tilde\mu,\mu_{ext})\Vert_\infty\leq K_i(\mu_{ext})\Vert \mu-\tilde\mu\Vert_\infty,\qquad \mu,\tilde\mu\in D_i(\mu_{ext})\subset S_A,\ i=1,\dots, n
\end{equation}
for domains that cover $S_A$. Since 
$\mathrm{Tr}\, F(\mu,\mu_{ext})=0$ $\mathrm{Tr}\,\mu(t)=1, t\geq 0$ will follow for integral curves. The domain 
$D_i(\mu_{ext})\subset S_A$ in (\ref{Lip_cont}) is defined as follows: 
For $\nu\in S_A^{int}$ let $C_i(\nu)$ be the $(n-1)$-dimensional  affine convex cone in $H_A$ 
with base point $\nu\in S_A^{int}$ and generating vectors 
$\nu- P_k,k=1,\dots, i-1,i+1,\dots, n$ being linearly independent if $\nu$ is in the interior of $S_A$. $C_i(\nu)$ lies in the one co-dimensional hyperplane in $H_A\simeq\RE^n$ that contains 
$S_A$. Let $D_i(\nu):= C_i(\nu)\cap S_A$. Clearly, 
$\cup_{i=1}^n D_i(\nu)=S_A$ and the intersections $D_{i_1}(\nu)\cap\dots\cap D_{i_m}(\nu), 2\leq m\leq n$ are the $(n-m)$-dimensional common boundaries of them. For the special case $n=3$ see Figure \ref{D_domains}. 

Let $\mu\in D_i(\mu_{ext})$ and let $\mu=\sum_k r_k P_k$ and $\mu_{ext}=\sum_k s_k P_k$ be the corresponding convex combinations. Since $\mu_{ext}\in S_A^{int}$ by assumption we have $s_k>0, k=1,\dots, n$. 
Since $\mu\in D_i(\mu_{ext})$ by assumption, that is $\mu_{ext}\in S_i(\mu)$, $\mu_{ext}$ can be written as the convex combination: $\mu_{ext}=\lambda_i\mu+\sum_{k\not= i} \lambda_kP_k$, which implies $r_i\geq s_i$.  
Hence, $\lambda\equiv\lambda(\mu,\mu_{ext})=\lambda_i=s_i/r_i$ in $f(\mu,\mu_{ext})$ given in (\ref{nonlin_eff_dyn_SM}) and 
\begin{equation}\label{tangent_vector_inthecone}
F(\mu,\mu_{ext})\equiv 
F\left(\sum_k r_k P_k,\sum_k s_k P_k\right)
=a\sum_{k\not= i} r_k\lambda_k(\mu- P_k)
=a\sum_{k\not= i} r_k(s_k-\frac{s_i}{r_i}r_k)(\mu- P_k)
\in C_i(\mu).
\end{equation}
Therefore if $\mu,\tilde\mu\in D_i(\mu_{ext})$ one can use the form (\ref{tangent_vector_inthecone}) of the tangent vectors to prove that (\ref{Lip_cont}) holds with Lipschitz constants $K_i(\mu_{ext})=a(4+6/s_i)$ in $D_i(\mu_{ext}), i=1,\dots, n$. To get these constants one uses triangle inequality for the norm, that $\Vert\ \Vert_\infty$ is bounded by $1$ on $S_A$,  non-negativity of the coefficients $r_k,\tilde r_k,s_k, k=1,\dots, n$, that they sum up to 1, the Cauchy--Schwarz inequality for their scalar product in $\RE^n$, and the inequality $r_i\geq s_i$. Thus, due to Lipschitz continuity, for any $\mu_{ext}\in S_A^{int}$ there exists a unique solution of the initial value problem of (\ref{nonlin_eff_dyn_SM}) with $\mu(0)\in D_i(\mu_{ext}), i=1,\dots, n$ till the integral curve remains in the corresponding domain $D_i(\mu_{ext})$.  

Due to (\ref{tangent_vector_inthecone}) the tangent vector at the point $\mu$ is in the `future' affine cone $C_i(\mu)$, which contains all cones $C_i(\tilde\mu)$ with $\tilde\mu\in D_i(\mu):=C_i(\mu)\cap S_A$. Therefore the unique integral curve with initial condition $\mu(0)\in D_i(\mu_{ext})$ gives rise to  monotone decreasing affine cones, $C_i(\mu(\tilde t)) \subseteq C_i(\mu(t))$ if $\tilde t\geq t$, which is strictly monotone at $t$ if $F(\mu(t),\mu_{ext})\not=0$. This property excludes closed integral curves.
Since $\mu(0)\in D_i(\mu_{ext}):=C_i(\mu_{ext})\cap S_A$, that is
$C_i(\mu(0)) \subseteq C_i(\mu_{ext})$, the unique integral curve $\mu(t)$ does not leave the affine cone $C_i(\mu_{ext})$. It does not leave 
$D_i(\mu_{ext})$, that is $S_A$, either, because $\mu(t)$ at the boundary face of $S_A$ characterized by a zero $P_k$ coefficient has tangent vector parallel with that face: due to (\ref{tangent_vector_inthecone}) the `normal' component $ar_k(t)\lambda_k(t)(\mu(t)-P_k)$ of the tangent vector $F(\mu(t),\mu_{ext})$ is zero because $r_k(t)=0$. 

Let us turn to the fixed point structure of the dynamics (\ref{nonlin_eff_dyn_SM}), that is to the possible asymptotic states. Since integral curves with initial points $\mu(0)\in D_i(\mu_{ext})$ do not leave $D_i(\mu_{ext})$ the integral curves with $\mu(0)\in D_{i_1}(\mu_{ext})\cap\dots\cap D_{i_m}(\mu_{ext}), 2\leq m\leq n$ do not leave the corresponding common boundary $D_{i_1}(\mu_{ext})\cap\dots\cap D_{i_m}(\mu_{ext})$. For $m=n$ the intersection contains the single point $\mu_{ext}$, hence, it is a fixed point in accordance with the fact that $F(\mu_{ext},\mu_{ext})=0$ due to $f(\mu_{ext},\mu_{ext})=0$. It is the only fixed point of (\ref{nonlin_eff_dyn_SM}) in $S_A^{int}$, because if $\mu=\sum_k r_kP_k\in D_i(\mu_{ext})$ with all $r_k>0$ and $\lambda_i <1$ then 
$F(\mu,\mu_{ext})\not=0$ due to (\ref{tangent_vector_inthecone}). 
Therefore the fixed points different from 
$\mu_{ext}$ are on the boundary $\partial S_A$ of $S_A$. 

If $\mu(0)\in D_i(\mu_{ext})^{int}$, that is if $\mu_{ext}\in S_i(\mu(0))^{int}$, then $\mu_{ext}\in S_i(\mu(t))^{int}$ for all $t\geq 0$, which implies $\lambda_k(t)>0, k=1,\dots, n$. 
Hence, due to the corresponding expression of $F(\mu,\mu_{ext})$ in (\ref{tangent_vector_inthecone}) such initial states necessarily lead to the asymptotic state characterized by $r_k=0, k\not= i$,  that is by $r_i=1$, which state is nothing else than the fixed point $P_i$ satisfying $F(P_i,\mu_{ext})=0$ due to (\ref{tangent_vector_inthecone}).
Thus all the other asymptotic states on the boundary $\partial S_A$  should arise from initial states in a common boundary $D_{i_1}(\mu_{ext})\cap\dots\cap D_{i_m}(\mu_{ext}), 2\leq m< n$. Since the corresponding integral curve remains in the intersection the possible asymptotic states are on $\partial S_A\cap D_{i_1}(\mu_{ext})\cap\dots\cap D_{i_m}(\mu_{ext})$.
Therefore the only stable fixed points are $P_i, i=1,\dots, n$, because they and only they have a neighborhood in $S_A$ as attractor regions. The unstable fixed points on 
$\partial S_A\cap D_{i_1}(\mu_{ext})\cap\dots\cap D_{i_m}(\mu_{ext})$ can be characterized by the $d < n-1$ dimension of their attracting submanifold in $S_A$, which is $n-m$ in the generic case. We do not solve the $\mu_{ext}$-dependent fixed point condition $F(\mu,\mu_{ext})=0$ for them.  
$\mu(0)=\mu_{ext}$ is a maximally unstable fixed point because any other initial state in the $\mu_{ext}$-neighborhood $S_A^{int}$ leads to a different fixed point as asymptotic state.

ii) Let the initial state $\mu(0)=\sum_k r_k(0)P_k$ be fixed in repeated runs. 
We have seen that if the dynamics (\ref{nonlin_eff_dyn_SM}) is characterized by $\mu_{ext}\in S_i(\mu(0)))^{int}$, that is 
$\mu(0)\in D_i(\mu_{ext})^{int}$, then the asymptotic state is $P_i$. 
Hence, ff the external density matrix $\mu_{ext}\in S_A$ is chosen uniformly random within $S_A$ then the probability (relative frequency) of the asymptotic state $P_i$ is just the relative volume $V(S_i(\mu(0))^{int})/V(S_A)$, that is the quotient of the Lebesgue measures of $S_i(\mu(0))^{int}$ and $S_A$. 
Since the $(n-1)$-dimensional simplices $S_i(\mu(0))$ and $S_A$ in the unit cube of $\RE^n$ have a common $(n-2)$-dimensional boundary face the ratio of their volumes is given by the ratio of the distances of the remaining vertices $\mu(0)\in S_i(\mu(0))$ and $P_i\in S_A$ from that common boundary face. The ratio of these distances is nothing else than $r_i(0)=\mathrm{Tr}\, \mu(0)P_i$.

The probability of the outcome of an unstable fixed point is zero, because such an outcome implies that $\mu_{ext}$ is contained in a common boundary $S_{i_1}(\mu(0))\cap\dots\cap S_{i_m}(\mu(0)), 2\leq m< n$, which is a subset of  zero Lebesgue measure in $S_A$.\qed

\section{Closing remarks}

Apart from the lack of derivation of the effective description of selective measurements our toy model has two `technical' weakness as well: it works in finite dimensional Hilbert spaces and it is a two-step dynamics, which requires two consecutive asymptotic evolutions of an initial state. These technical weakness can be relieved a bit.

If both the Hilbert space $\HS$ both the abelian von Neumann algebra $\langle A\rangle\subset\BO(\HS)$ affiliated to a (possibly unbounded) self-adjoint observable $A$ are infinite dimensional then it is not clear how to choose the Lindblad operators to get $A$-decohered asymptotic states from any normal initial states on 
$\BO(\HS)$ as a result of a Lindblad dynamics. However, once an 
$A$-decohered asymptotic state is reached the second non-linear dynamical step can be used as an approximation: One 
can prescribe a finite partition of the spectrum $\sigma(A)\subseteq\RE$ of $A$, hence the corresponding spectral interval   projections $P_i\in\langle A\rangle, i=1,\dots, n$ generate a finite dimensional unital abelian algebra $\OA$ in 
$\langle A\rangle\subset\BO(\HS)$. Then the restriction of the original normal state on $\BO(\HS)$ to $\OA$, which should be equal to the restriction of the asymptotic $A$-decohered state of a Lindblad dynamics, can be the initial state of the second step non-linear dynamics described in the previous chapter.

One can also incorporate the simultaneous measurement (of the joint spectrum) of commuting observables $A, B, C\dots $ in the toy model. Since they generate an abelian algebra one can use the products of their spectral projections as minimal projections if they have finite spectra, otherwise products of the above mentioned spectral interval projections can be used as a certain approximation of the outcomes. Thus one can have, for example, an effective finite dimensional description of position measurements of commuting coordinate operators with an arbitrary fine but finite spectral, i.e. position resolution. 

The idealized two-step description of the selective measurement of $A\in M_n$, i.e. the use of Lindblad dynamics (\ref{Lindblad_equation}) followed by the non-linear dynamics (\ref{nonlin_eff_dyn_SM}), could be cured by combining them into a single dynamics on the set $S_n$ of density matrices in $M_n$: 
\begin{equation}\label{combined_evolution}
\frac{d\rho}{dt}=\hat L(\rho)+ F(\rho,\mu_{ext}),\quad 
\rho\in S_n,\mu_{ext}\in S_A,
\end{equation}
where the operators $H,V_k$ in the Lindblad generator 
$\hat L=\hat L(H,V_k)$ is chosen according to Proposition \ref{A-decoherence} to ensure $A$-decoherence. 
The tangent vector 
$F(\rho,\mu_{ext}):=\tilde f(\rho,\mu_{ext})-
\rho\,\mathrm{Tr}\, \tilde f(\rho,\mu_{ext})$ can be of the form given in (\ref{nonlin_eff_dyn_SM}), but  
$\tilde f\colon S_n\times S_A\to M_n$ should be an extension of
$f\colon S_A\times S_A\to\langle A\rangle$ with self-adjoint images. For example, one can choose 
$\tilde f(\rho,\mu_{ext}):=a(\lambda\rho^2-\sqrt{\mu_{ext}} \rho\sqrt{\mu_{ext}})$ with the extended definition of $\lambda\equiv\lambda(\rho,\mu_{ext})$: it is the maximal value of $\kappa\in[0,1]$ for which $\mu_{ext}-\kappa\rho$ is a positive operator in $M_n$. We think that if the parameter $a>0$ in $\tilde f$ is small enough, that is the $A$-decoherence is much faster than 
$A$-purification, then (\ref{combined_evolution}) leads to an effective two-step dynamics with results described in the previous chapter.

Finally, let us make some remarks about a possible experimental test of the dynamical nature of selective measurements, more precisely, about the test of the effective non-linear dynamical step in our toy model. Here the repeated measurements of $A$ with identical initial state $\mu(0)\in S_A$ can be thought as a $\mu(0)$-dependent map from probability distributions on external density matrices $\mu_{ext}\in S_A$ into probability distributions on the set $\{ P_1,\dots, P_n\}\subset S_A$ of asymptotic states. If the measuring device can be switched off and on quickly enough at $t>0$ without disturbing the intermediate state $\mu(t)$ of the measured system then one obtains a $\mu(0)$-dependent distribution of the intermediate states at time $t$ on the one hand, on the other hand the `immediately' restarted experiments with this intermediate state distribution as an initial state distribution lead to a $\mu(t)$-dependent distribution of asymptotic states 
$\{ P_1,\dots, P_n\}$, which is different from the asymptotic state distribution of uninterrupted measurements in general. 
The resulted asymptotic state distribution of repeated interrupted and restarted experiments with identical initial states 
$\mu(0)$ and identical interruption times $t$ can be calculated from the effective toy dynamics in principle, and can be compared with the relative frequency of the asymptotic experimental outcomes for any interruption time $t>0$.  
 
\emph{Acknowledgements.} The author is grateful to László B. Szabados for the fruitful discussions on first order differential equations. Special thanks to Lajos Diósi and László B. Szabados
for the careful reading of the manuscript and for the suggestions to improve it. 
Thia work has been supported by the Hungarian Scientific Research Fund, OTKA K-108384.

\end{document}